\documentclass[aps, superscriptaddress, reprint]{revtex4-1}

\usepackage{graphicx}
\usepackage{bm}
\usepackage{dcolumn}
\usepackage{amssymb,amsmath}
\usepackage[colorlinks=true,allcolors=blue]{hyperref}
\usepackage{color}
\usepackage{siunitx}

\usepackage{lipsum}
\newcommand{\cOut}[1]{}
\newcommand{\figref}[2]{\hyperref[#1]{\ref{#1}(#2)}}

\begin{document}

\title{Piezostrain -- a local handle to control gyrotropic dynamics in magnetic vortices}

\author{Vadym Iurchuk}
\email[e-mail: ]{v.iurchuk@hzdr.de}
\affiliation{Institute of Ion Beam Physics and Materials Research, Helmholtz-Zentrum Dresden-Rossendorf, 01328 Dresden, Germany}

\author{Serhii Sorokin}
\affiliation{Institute of Ion Beam Physics and Materials Research, Helmholtz-Zentrum Dresden-Rossendorf, 01328 Dresden, Germany}

\author{J\"urgen Lindner}
\affiliation{Institute of Ion Beam Physics and Materials Research, Helmholtz-Zentrum Dresden-Rossendorf, 01328 Dresden, Germany}

\author{J\"urgen Fassbender}
\affiliation{Institute of Ion Beam Physics and Materials Research, Helmholtz-Zentrum Dresden-Rossendorf, 01328 Dresden, Germany}
\affiliation{Institute of Solid State and Materials Physics, Technische Universit\"at Dresden, 01062 Dresden, Germany}

\author{Attila K\'akay}
\affiliation{Institute of Ion Beam Physics and Materials Research, Helmholtz-Zentrum Dresden-Rossendorf, 01328 Dresden, Germany}

\date{\today}

\begin{abstract}
We present a study of the piezostrain-tunable gyrotropic dynamics in Co$_{40}$Fe$_{40}$B$_{20}$ vortex microstructures fabricated on a 0.7Pb[Mg$_{1/3}$Nb$_{2/3}$)]O$_3$–0.3PbTiO$_3$ single crystalline substrate. Using field-modulated spin rectification measurements, we demonstrate large frequency tunability (up to 45~\%) in individual microdisks accessed locally with low surface voltages, and magnetoresistive readout. With increased voltage applied to the substrate, we observe a gradual decrease of the vortex core gyrotropic frequency associated with the strain-induced magnetoelastic energy contribution. The frequency tunability strongly depends on the disk size, with increased frequency downshift for the disks with larger diameter. Micromagnetic simulations suggest that the observed size effects originate from the joint action of the strain-induced magnetoelastic and demagnetizing energies in large magnetic disks. These results enable a selective energy-efficient tuning of the vortex gyrotropic frequency in individual vortex-based oscillators with all-electrical operation.
\end{abstract}

\maketitle


\section{Introduction}
Stable topological objects can be spontaneously formed in confined micro- and nanostructures with high symmetry, e.g. squares, disks and ellipses, as a result of a competition between exchange and magnetostatic energies~\cite{aharoni_upper_1990, usov_magnetization_1993}. These topologically protected magnetic states -- \textit{magnetic vortices} -- are characterized by a curling in-plane magnetization an out-of-plane singularity in the center of the structure, known as the vortex core (VC)~\cite{shinjo_magnetic_2000}.
Resonant excitation of the magnetic vortex by Oersted field and/or spin-polarized electrical current results in a dynamical gyrotropic motion of the VC at a specific frequency~\cite{guslienko_eigenfrequencies_2002}, usually in sub-GHz range. Resonant VC gyration is successfully employed in vortex-based spin-torque oscillators, thanks to the reduced dynamical noise of the VC gyrotropic mode~\cite{lebrun_nonlinear_2014,lebrun_understanding_2015,litvinenko_2020,litvinenko_analog_2021}. However, one of the major drawbacks of the vortex-based oscillators is the low tunability of the gyrotropic frequency in the linear regime as a result of the topologically protected dynamical mode of the vortex core. Indeed, the gyration frequency, in the first approximation, is determined by the magnetic properties and the geometry of the oscillating layer~\cite{guslienko_eigenfrequencies_2002}. To overcome this limitation, different approaches have been attempted, including tuning by bias dc current and/or out-of-plane magnetic field~\cite{dussaux_large_2010}, bias in-plane field and/or rf excitation amplitude~\cite{ramasubramanian_effects_2022}, gyration frequency modification in magnetostatically coupled vortices~\cite{sluka_spin-torque-induced_2015}, or using focused-ion-beam-assisted modification of the intrinsic magnetic material parameters~\cite{ramasubramanian_tunable_2020}

An alternative way employs manipulation of the magnetic vortex configuration by means of static electric fields in composite piezoelectric/ferromagnetic heterostructures via strain-mediated magnetoelectric coupling. This coupling is based on the joint action of two effects: piezoelectric and magnetoelastic (inverse magnetostrictive). Strain-mediated methods of magnetization control attracted recently an increased attention due to reduced energy consumption and possibility of an indirect control, avoiding large currents through the magnetic element and/or local magnetic field application~\cite{roy_energy_2012,roy_ultra-low-energy_2013,iurchuk_multistate_2014,iurchuk_electrical_2015,wang_strain-mediated_2017,schneider_rf_2019,iurchuk_strain-controlled_2023}. When applied to the magnetic vortices, recent studies demonstrated the possibility to control the vortex configuration~\cite{parkes_voltage_2014,gilbert_magnetic_2016,ghidini_voltage-driven_2020,ghidini_voltage-driven_2020-1} or even to switch the direction of the vortex circulation or polarization~\cite{liElectricalSwitchingMagnetic2017, ostlerStrainInducedVortex2015} in magnetostrictive microdisks grown on piezoelectric substrates upon application of an electric field. Theoretical studies show that the vortex gyrotropic mode can also be modified by introducing an in-plane magnetoelastic anisotropy, resulting in a decrease of the gyration frequency due to the softening of the restoring force spring constants~\cite{roy_2013}. Experimentally, the magnetoelastic anisotropy-induced modification of the vortex gyration frequency was shown in~\cite{finizio_2017}, using time-resolved scanning transmission X-ray microscopy of the rf Oersted field-driven vortex dynamics in CoFeB microsquares on mechanically stretched Si$_3$N$_4$ membranes. More recently, piezoelectrical control over the rf field-driven vortex gyration trajectories was demonstrated by performing time-resolved photoemission electron microscopy combined with X-ray magnetic circular dichroism experiments on sub-micron sized Ni vortices under strain generated electrically in piezoelectric PMN-PT substrate~\cite{filianina_2019}.

One of the major obstacles of utilizing such approaches for spintronic device prototypes is the non-local nature of the electrical excitation of the piezosubstrate. Usually, it requires large voltages applied across the substrate thickness, generating bulk strains in an entire piezosubstrate, thus hindering the selective access and control over individual devices fabricated on a single chip. In addition, a simple electrical detection of the vortex static and/or dynamic behavior, in contrast to cumbersome X-ray-based imaging methods, would facilitate the path towards implementation of strain-tunable spintronic oscillators.

Here, we present a study of the piezostrain-tunable vortex core gyrotropic dynamics in Co$_{40}$Fe$_{40}$B$_{20}$ (hereafter CoFeB) circular microstructures grown on piezoelectric 0.7Pb[Mg$_{1/3}$Nb$_{2/3}$)]O$_3$–0.3PbTiO$_3$ (hereafter PMN-PT) substrates. Using spin rectification measurements, we demonstrate large gyrotropic frequency tunability (up to 45~\%) in individual disks accessed locally with low surface voltages ($\leqslant$16~V), and all-electrical operation. With increased voltage applied to the PMN-PT, we observe gradual decrease of the VC gyrotropic frequency associated to the strain-induced magnetoelastic energy contribution due to the inverse magnetostrictive (magnetoelastic) effect. Moreover, the frequency tunability strongly depends on the magnetic disk size, with increased frequency downshift for the disks with larger diameter. By analyzing the simulated strain-dependent energies for different disk sizes, we attribute the observed size effects to the joint action of the strain-induced magnetoelastic and demagnetizing energies in large magnetic disks.


\section{Details on sample preparation, experimental setup and micromagnetic simulations}
We use (011)-cut PMN-PT single crystals as functional piezoelectric substrates capable of generating high strains upon electric field application~\cite{zhang_2008, wu_2011}.
Surface electrodes, magnetic microdisks and the contact pads were fabricated on the PMN-PT substrates in a three-step lithography process. One has to note, that for the PMN-PT compounds near the morphotropic phase boundary, the crystallographic rhombohedral-tetragonal phase transition occurs at $\sim$90$^{\circ}$C and the ferroelectric Curie temperature is $\sim$140$^{\circ}$C~\cite{guo_2002, guo_2003, makhort_2018}. Therefore, conventional lithography processes, which include high-temperature pre- and/or postbaking of the photo- and e-beam resists spun on the substrates, may induce the irreversible crystallographic phase transitions in the PMN-PT. This can potentially lower its piezoelectric properties due to residual stresses in the crystal or even lead to formation of cracks on the surface of the crystal, thus making it unsuitable for the microfabrication of thin film devices. To avoid overheating of the PMN-PT substrate, we developed a specific low-heat three-step fabrication process for the fabrication of the surface electrodes, magnetic microdisks and contact pads. First, the surface electrodes to generate a local strain in the PMN-PT substrate, were fabricated by UV lithorgaphy, e-beam metallization with Cr(5~nm)/Au(125~nm) and conventional lift-off. As the next step, the magnetostrictive microdisks were patterned by means of electron beam lithography, followed by magnetron sputtering of a Cr(5~nm)/CoFeB(30~nm)/Cr(2~nm) film and a lift-off. The bottom and top Cr layers were used as seed and cap layers, respectively. The disk diameters were chosen to fulfill the geometric criterion of the vortex formation in ferromagnetic disks~\cite{jubert_analytical_2004}. As the final step, the contact pads were fabricated by electron beam lithography, e-beam evaporation of Cr(5~nm)/Au(50~nm) and lift off to provide individual electrical access to each microdisk.

\begin{figure*}[t]
\centering
    \includegraphics[width=\textwidth]{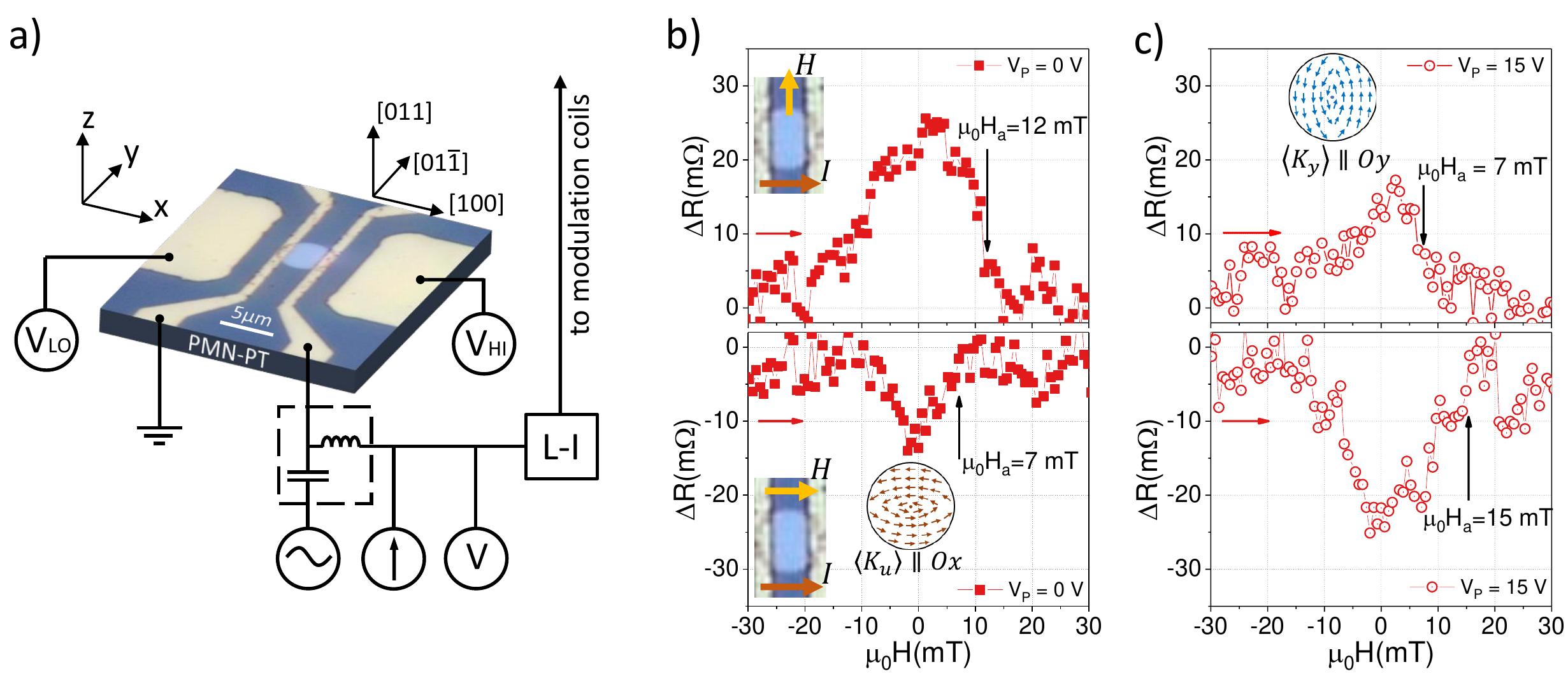}
    \caption{(a) Schematics of the experiment enabling a detection of the magnetization dynamics through a rectified dc voltage measurements via lock-in technique, with a simultaneous application of an electric field to the PMN-PT substrate. (b) Magnetoresistance of the CoFeB disk (diameter $d$ = 3.65~$\mu$m) measured at $I_{dc}$ = 2~mA and $V_P$ = 0~V for the magnetic field $H$ applied perpendicular (top graph) and parallel (bottom graph) to the dc current direction. The magnetic field sweeping direction is indicated by the red arrow. The insets show the device images with the corresponding $I_{dc}$ and $H$ directions marked with arrows. (c) Magnetoresistance of the same device as shown in panel (b) measured at $I_{dc}$ = 2~mA and $V_P$ = 15~V for $H \perp I_{dc}$ (top graph) and $H \parallel I_{dc}$ (bottom graph). Presumable configurations of the magnetic vortex deduced from the magnetoresistance measurements are schematically depicted in the corresponding insets of panels (a) and (b) with red (blue) arrows showing the local direction of the magnetic moments within the disk.} 
    \label{fig1}
\end{figure*}

To detect the VC dynamics, we used a standard magnetotransport setup with rf capability [see Fig.~\ref{fig1}(panel a)] for electrical detection of magnetization dynamics in single magnetic vortices at room temperature. The detection technique exploits the anisotropic magnetoresistance (AMR) effect, i.e., the resistance change induced by the relative angle between the direction of the electrical current and the net magnetization of a magnetic structure. An rf current injected through a bias-T into the microdisk device excites the VC gyrotropic dynamical mode, via the joint action of the spin-transfer torque and rf Oersted field, and thereby leads to a dynamical magnetoresistance oscillating at the excitation frequency. The time-averaged product of the rf current and the dynamical magnetoresistance –- which results in a rectified dc voltage $V_{dc}$ --– is measured by a conventional homodyne detection scheme using a lock-in amplifier. When the excitation frequency matches the eigenfrequency of the gyrotropic mode, the resulting $V_{dc}$ is enhanced due to the dynamical magnetoresistance increase associated with the resonant expansion of the VC gyration trajectory. To improve the signal-to-noise ratio, magnetic field modulation of the dynamical magnetoresistance at the lock-in reference frequency (here 1033~Hz) was used similar to~\cite{ramasubramanian_effects_2022, ramasubramanian_thesis_2022}.
To allow for an electrical excitation of the piezoelectric PMN-PT substrate, a dc voltage $V_P = V_{HI} - V_{LO}$ is applied between the surface electrodes, as depicted in Fig.~\ref{fig1}(panel a).

For the micromagnetic simulations, we use the GPU-accelerated \textsc{MuMax3} software package~\cite{vansteenkiste_2014} to simulate the magnetization dynamics in the CoFeB vortices under strain. Similar to~\cite{wagner_2021,iurchukStressinducedModificationGyration2021b}, the VC dynamics is excited by an in-plane rf magnetic field pulse $b_{rf} \textit{sinc} (2 \pi f_{cut} t) $ with the amplitude $b_{rf}$ = 2.5~mT and the cut-off frequency $f_{cut}$ = 2~GHz. For each simulation, the time evolution of the magnetization is recorded for 200~ns with the time step of 50~ps, and the microwave-absorption power spectra are calculated by performing the Fourier transform of the time-dependent magnetization dynamics.
We use the following CoFeB material parameters: saturation magnetization $M_s$ = 1100~kA/m (measured by vibrating sample magnetometry), exchange constant $A_{ex}$ = 20~pJ/m$^3$, and damping parameter $\alpha$ = 0.008. For each disk diameter, we use 5$\times$5$\times$30~nm$^3$ cell sizes for the magnetization dynamics simulations, and a finer discretization into 5$\times$5$\times$5~nm$^3$ cells, for the computation of the equilibrium energies.

\section{Results and discussion}
\subsection{Strain-dependent magnetoresistance} \label{strain-dep MR}
Fig.~\ref{fig1}(panel b) shows the typical anisotropic magnetoresistance of the CoFeB disk with a diameter of 3.65~$\mu$m measured for a dc current $I_{dc}$ = 2~mA and at zero $V_P$ voltage applied to the PMN-PT. Top graph of Fig.~\ref{fig1}(panel b) shows the magnetoresistance (MR) curve for the magnetic field $H_{\perp}$ applied perpendicular to the $I_{dc}$ direction. Bottom graph in Fig.~\ref{fig1}(panel b) shows the MR for $H_{\parallel}$, i.e. for $H \parallel I_{dc}$. The magnetic field is swept from negative to positive saturation. For $H \perp I_{dc}$, when the field is increased from the negative saturation values ($\approx$40mT) towards zero, we observe a resistance growth of $\approx$25~m$\Omega$ (corresponding to the MR$_{\perp}$ ratio of $\approx$0.015~\% being a typical value for Fe-based alloys). This resistance increase indicates a nucleation of a vortex within the CoFeB disk. Further increase of the magnetic field leads to the VC shift towards the edge of the disk and, eventually, to the VC expulsion when the annihilation field $H_a$ is reached. Similarly, for $H \parallel I_{dc}$, the resistance drop of $\approx$15~m$\Omega$ (MR$_{\parallel} \approx$0.008\%) is measured in the vicinity of $H_{\parallel}$=0, in agreement with the angular dependence of MR.

However, two notable observations can be made. First, a large asymmetry of the MR magnitudes for $H \parallel I_{dc}$ and $H \perp I_{dc}$, where the value of MR$_{\perp}$ is almost twice higher as compared to MR$_{\parallel}$. Second, a large difference in the vortex nucleation/annihilation fields is present, depending on the magnetic field orientation. From the magnetoresisance data, we estimate the annihilation field values $\mu_0 H_{a_\perp}$ = 12~mT and $\mu_0 H_{a_\parallel}$ = 8~mT. This indicates an asymmetric magnetic configuration of the nucleated vortex, otherwise, in case of a radially symmetric vortex, a rather similar magnetoresistive response would be expected for all in-plane directions of the applied field. The distortion of the vortex configuration is attributed to the presence of the uniaxial magnetic anisotropy, typical for the sputtered CoFeB films~\cite{gladii_2023}. In our measurements, MR$_{\parallel} <$ MR$_{\perp}$ and $H_{a_\parallel} < H_{a_\perp}$, which indicates the presence of a non-zero net magnetic anisotropy $\langle K_u \rangle$ along the $H_{\parallel}$, i.e. along the $x$ direction. The corresponding vortex configuration is sketched in the inset of the Fig.~\ref{fig1}(panel b). We estimate the anisotropy constant $K_{u_x}$ from the difference between the annihilation fields $H_{a_\parallel}$ and $H_{a_\perp}$ as $\langle K_{u_x} \rangle = \left( H_{a_\parallel} - H_{a_\perp} \right) / 4 M_s $. Using the measured values of $H_{a_\parallel}$ and $H_{a_\perp}$ and the $M_s$ = 1110~kA/m (from the magnetometry measurements), we obtain $K_{u_x} \approx$ 1.1~kJ/m$^3$.

Fig.~\ref{fig1}(panel c) shows the magnetoresistance curves of the same disk measured for the $I_{dc}$=2~mA, under $V_P$ = 15~V applied to the PMN-PT. We note a drastic qualitative difference between the MR curves measured with and without $V_P$. Comparing the MR curves measured for $H \perp I_{dc}$, at $V_P$ = 0 and $V_P$ = 15~V [see top graphs in Fig.~\ref{fig1}(b) and (c)], we observe two main effects, namely a reduction of the MR$_{\perp}$ ratio from 0.015\% to 0.01\% and a decrease of the $H_{a_\perp}$ from 12 to 7~mT . On the other hand, for $H \parallel I_{dc}$, an opposite effect is observed, i.e. increase of both MR$_{\parallel}$ (from 0.008\% to 0.015\%) and $H_{a_\parallel}$ (from 8 to 15~mT).

\begin{figure}[t]
\centering
    \includegraphics[width=0.48\textwidth]{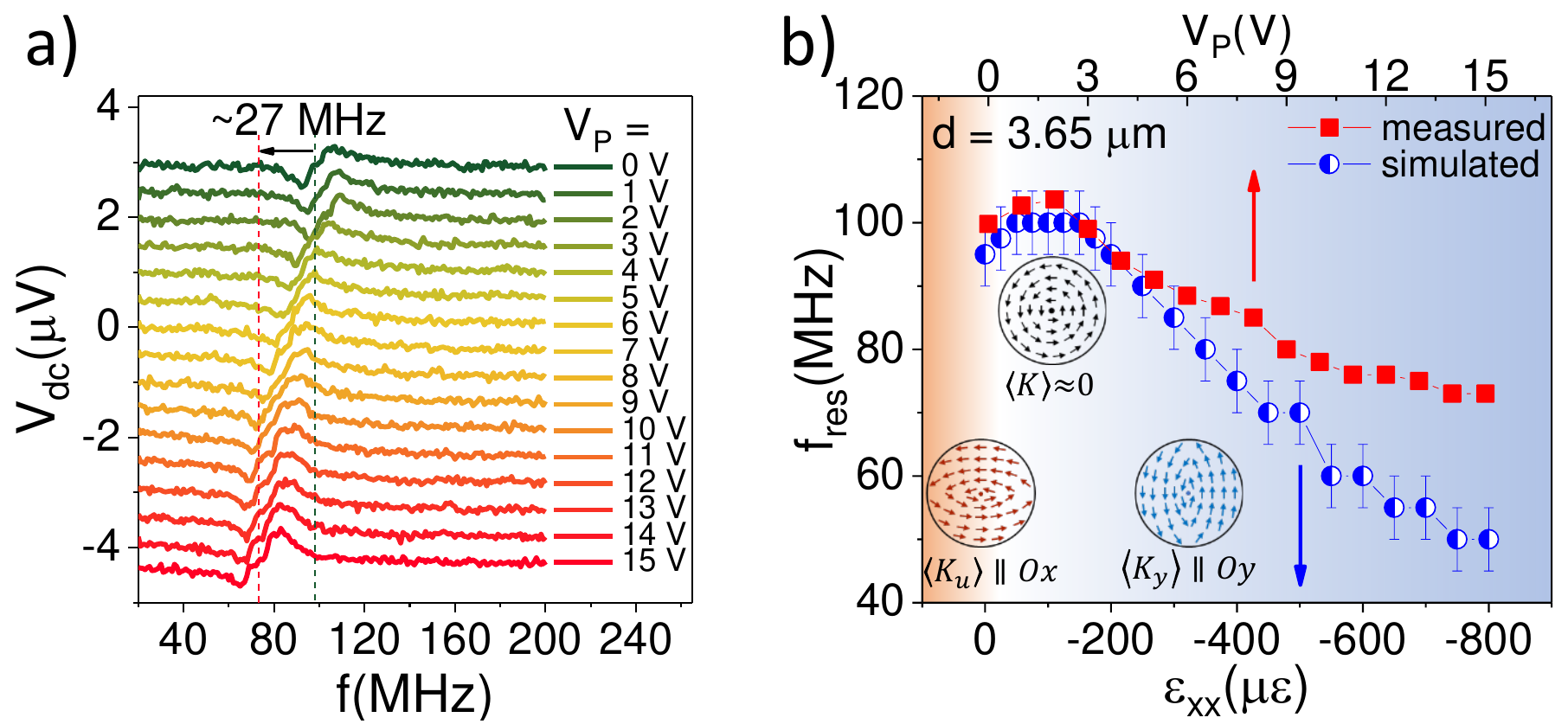}
    \caption{(a) Frequency-swept zero-field rectification spectra of the 30-nm-thick CoFeB disk (diameter 3.65 µm) for the different $V_P $voltages applied to the PMN-PT. (b) Red squares: VC gyrotropic frequency $f_0$ vs. $V_P$ extracted from (a). Blue circles: simulated values of the gyrotropic frequency vs. uniaxial strain $\varepsilon_{xx}$. For the micromagnetic simulations, the uniaxial anisotropy $K_u$ = 1~kJ/m$^3$ $\parallel Ox$ was introduced. The sketches show the schematic vortex configurations in the anisotropy compensation region ($\langle K_u \rangle \parallel Ox$), zero anisotropy region ($\langle K_u \rangle \approx 0$) and anisotropy enhancement region ($\langle K_u \rangle \parallel Oy$), respectively (see text for details).}
    \label{fig2}
\end{figure}

These voltage-induced effects are explained by the presence of a magnetoelastic anisotropy energy $W_{\varepsilon}$ due to the electric-field induced strain, generated via converse piezoelectric effect in PMN-PT and transferred to the CoFeB microdisk. The magnetoelastic energy density is defined as $K_{\varepsilon} = \frac{3}{2} \lambda_s Y \langle \varepsilon_{xx} \rangle$, where $\lambda_s$ = 50~ppm is the saturation magnetostriction of CoFeB, $Y$ = 160~GPa is the CoFeB Young's modulus and $\langle \varepsilon_{xx} \rangle$ is the net strain along the $x$ direction. We introduce $\langle \varepsilon_{xx} \rangle$ as effective uniaxial strain generated locally in the small area between the surface electrodes (see Fig.~\ref{fig1}(panel a)). Based on the magnetoresistance measurements, we determine that $\langle \varepsilon_{xx} \rangle < 0$, i.e. the generated strain is compressive along the $x$ direction. Indeed, the observed increase of the MR$_{\parallel}$ and $H_{a_\parallel}$ [see bottom graph in Fig.~\ref{fig1}(panel c)] suggests that, under increasing strain, the net anisotropy $\langle K_{u_x} \rangle$ decreases, in agreement with $\langle \varepsilon_{xx} \rangle < 0$. Under compressive uniaxial strain, the magnetic vortex configuration is distorted, which is manifested as the contraction in the direction of the strain and the elongation in the direction perpendicular to the strain. Therefore, the compressive strain along the $x$ direction lowers the effective anisotropy in this direction and simultaneously increases the net anisotropy in the perpendicular in-plane direction, i.e along $Oy$. This is consistent with the measured reduction of the MR$_{\perp}$ and $H_{a_\perp}$ at $V_P$ = 15~V, and for the magnetic field perpendicular to the current. The vortex configuration corresponding to the strained state is schematically shown in the inset of Fig.~\ref{fig1}(panel c).
One has to note that, in our experiment, the electric field is applied along [100] crystallographic axis of the PMN-PT crystal and, therefore, the compressive strain is expected for a given crystallographic orientation of the PMN-PT crystal~\cite{wu_2011}.  

\begin{figure}[b]
    \includegraphics[width=0.48\textwidth]{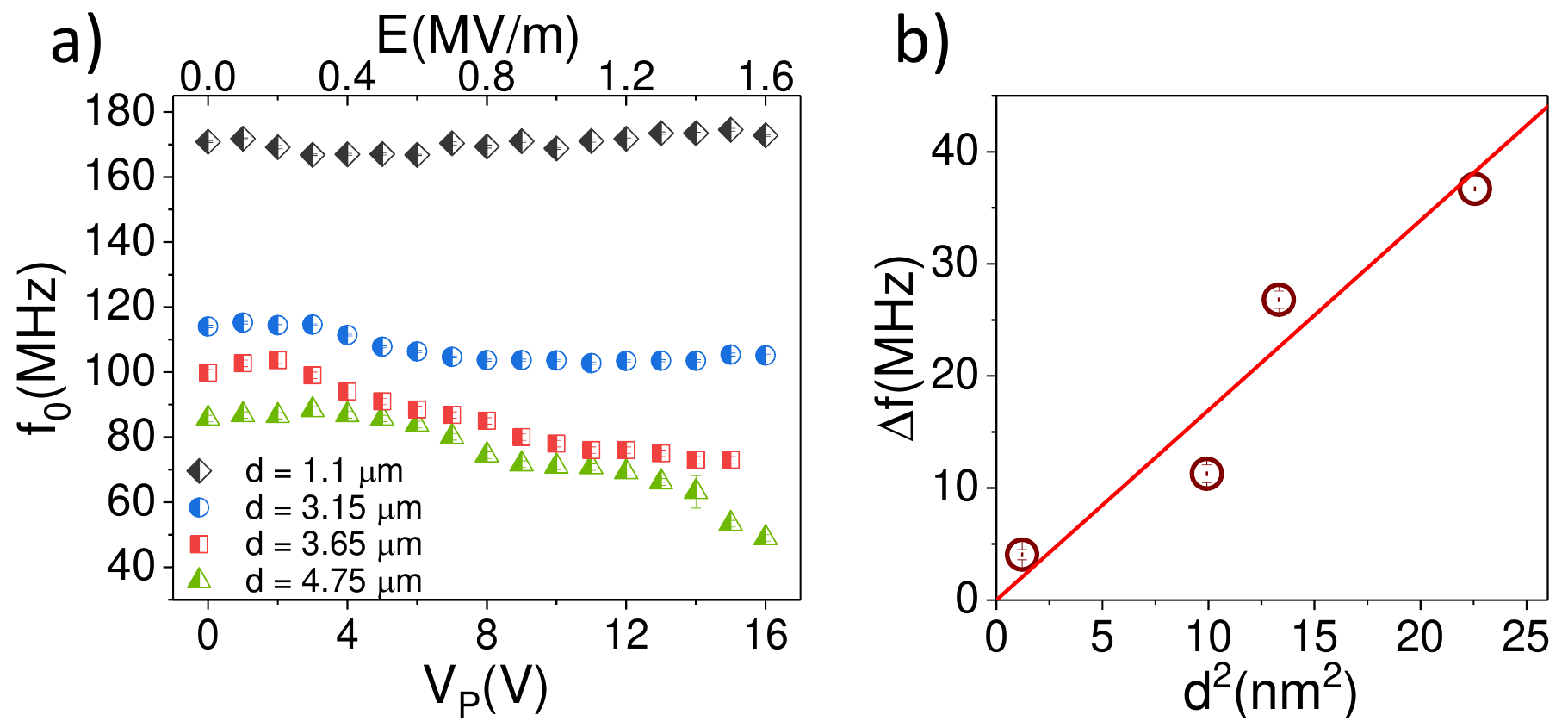}
    \caption{(a) Gyrotropic frequency vs. applied voltage (electric field) for different device diameters $d$: 1.1~$\mu$m (black diamonds); 3.15~$\mu$m (blue circles); 3.65~$\mu$m (red squares) and 4.75~$\mu$m (green triangles). (b) $\Delta f$ vs. $d^2$ extracted from the $f_0(V_P)$ dependencies. Red line is a linear fit to $\Delta f = const \cdot d^2$.}
    \label{fig3}
\end{figure}

\subsection{Strain-controlled gyration dynamics} \label{strain-dep gyration}

Fig.~\ref{fig2}(a) shows the rectified $V_{dc}$ spectra vs. rf current frequency for the CoFeB disk with 3.65 $\mu$m diameter, measured at zero magnetic field and for different values of the voltage $V_P$ applied to the PMN-PT substrate. For all $V_P$ values, the measured resonances have a typical antisymmetric Lorentzian shape, associated with the dominant contribution of the rf Oersted field to the VC gyration dynamics~\cite{kimDoubleResonanceResponse2013, ramasubramanian_effects_2022}. For $V_P$ = 0, $f_0$ = 100~MHz, in a good agreement with the analytically predicted value (112~MHz) by the Guslienko's "two-vortices" model~\cite{guslienko_eigenfrequencies_2002}. We observe a gradual decrease of the gyrotropic frequency with increasing $V_P$ (i.e. with increasing piezostrain) attributed to the magnetoelastic anisotropy-induced softening of the restoring force spring constants~\cite{roy_2013}. For the voltage range used in our experiment ($V_P \leqslant$ 15~V), the maximum frequency downshift is 27~MHz (corresponding to 27~\%) at 15~V, as compared to the initial value at 0~V.

\begin{figure*}[t]
    \includegraphics[width=\textwidth]{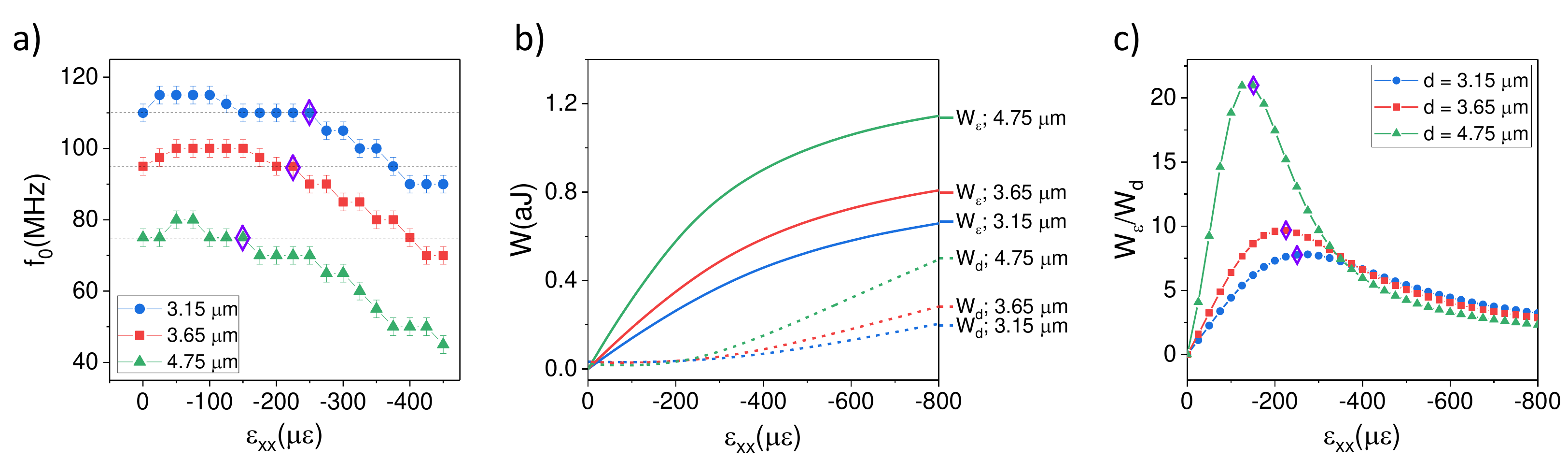}
    \caption{(a) Simulated $f_0(\varepsilon_{xx})$ for disk diameters of 3.15~$\mu$m (blue circles); 3.65~$\mu$m (red squares) and 4.75~$\mu$m (green triangles). For the given diameter, dotted lines indicate the $f_0$ level and diamonds mark the onset of the frequency decrease. (b) Calculated values of the magnetoelastic $W_{\varepsilon}$ (solid lines) and demagnetizing $W_d$ (dashed lines) energies as a function of strain for $d$ = 3.15~$\mu$m; 3.65~$\mu$m and 4.75~$\mu$m. (c) $W_{\varepsilon} / W_d$ ratios vs. $\varepsilon_{xx}$ for different $d$. Diamonds mark the onset of the frequency decrease observed in (a).}
    \label{fig4}
\end{figure*}

Fig.~\ref{fig2}(b, red squares) shows the values of the gyrotropic frequency $f_0$ as a function of the $V_P$. A detailed examination of the $f_0 (V_P)$ dependence reveals a non-monotonous behavior of the gyrotropic frequency, namely a small increase at low voltages ($V_P \leqslant$ 2~V), followed by a steady decrease for $V_P \geqslant$ 3~V. To understand this behavior, we conducted micromagnetic simulations of the VC gyration dynamics as a function of the uniaxial strain $\varepsilon_{xx}$. The corresponding $f_0(\varepsilon_{xx})$ dependence is plotted in Fig.~\ref{fig2}(b, blue circles) and is in good agreement with the experimental $f_0(V_P)$ for moderate values of the voltage $V_P \leqslant$ 6~V, corresponding to the uniaxial compressive strain $|\varepsilon_{xx}| \leqslant$ 350~$\mu \varepsilon$. The comparison between the experimental $f_0(V_P)$ and simulated $f_0(\varepsilon_{xx})$ allows for an estimation of the piezoelectrically generated strain per unit electric field E in the PMN-PT in the linear regime. We obtain $\varepsilon / E \approx$~--600 pm/V, in agreement with the typical values of the in-plane piezoelectric constants of the PMN-PT.
An increased discrepancy between the experimental $f_0(V_P)$ and simulated $f_0(\varepsilon_{xx})$ at high $V_P$ is related to the non-linear strain-voltage dependence in voltage range close to the ferroelectric saturation.

In Fig.~\ref{fig2}(b, blue circles), three distinctive regions in the $f_0(\varepsilon_{xx})$ dependence can be differentiated:
$|\varepsilon_{xx}| \leqslant$~75~$\mu \varepsilon$, where $f_0$ increases;
100~$\leqslant |\varepsilon_{xx}| \leqslant$~150~$\mu \varepsilon$, where $f_0$ is stable; and $|\varepsilon_{xx}| \geqslant$~175~$\mu \varepsilon$, where $f_0$ decreases. Taking into account the net uniaxial anisotropy $K_{u_x}$=1.1~kJ/m$^3$ in the as-prepared CoFeB disk (see section~\ref{strain-dep MR}), these three regions can be associated with the \textit{anisotropy compensation} region, where $\langle K_u \rangle \parallel Ox$; the \textit{zero anisotropy} region, where $\langle K_u \rangle \approx 0$; and the \textit{anisotropy enhancement} region, where $\langle K_u \rangle \parallel Oy$.
Indeed, when the compressive strain $\varepsilon_{xx}$ increases from 0 to $\approx$~75~$\mu \varepsilon$, the induced magnetoelastic anisotropy is aligned along the $y$ direction (i.e. perpendicular to the $x$ direction), and competes with the existing net anisotropy along $Ox$ (see the bottom left sketch in Fig.~\ref{fig2}(b)). The shape of the vortex is, therefore, gradually modified towards the radially symmetric configuration. This manifests as the increase of the gyrotropic frequency due to the decrease of the average uniaxial anisotropy.
When the strain-induced magnetoelastic anisotropy compensates the intrinsic anisotropy, the net in-plane anisotropy vanishes and the gyrotropic frequency reaches its maximum value, which corresponds to the gyrotropic frequency of the radially symmetric vortex (see the middle sketch in Fig.~\ref{fig2}(b)). Further increase of the $|\varepsilon_{xx}|$ value leads to the enhancement of the net anisotropy along $Oy$ (see the bottom right sketch in Fig.~\ref{fig2}(b)), accompanied by the reduction of the gyrotropic frequency $f_0$.

Our measurements demonstrate an efficient way to achieve a significant modification of the vortex gyrotropic frequency by piezoelectric strains, generated locally with moderate voltages applied to the piezosubstrate.

\subsection{Size-dependent frequency downshift. Role of the magnetoelastic and demagnetizing energies} \label{size-depencence}

Fig.~\ref{fig3}(a) shows the gyrotropic frequency $f_0$ vs. voltage $V_P$, measured in the CoFeB disks with different diameters $d$: 4.75, 3.65, 3.15, and 1.1~$\mu$m. Together with the diameter-dependent gyrotropic frequency $f_0$, we observe a clear dependence of the maximum frequency downshift $\Delta f$ on the device size. Here, $\Delta f$ is defined, for each disk diameter, as a difference between the gyrotropic frequency $f_0$ at $V_P = 0$ and the minimum attainable $f_0$ under strain (at $V_P > 0$). For $d$ = 1.1~$\mu$m, the gyrotropic frequency $f_0$ weakly depends on the voltage $V_P$ showing a decrease from 171~MHz at zero strain to 166~MHz at $V_P$ = 5~V followed by a slight increase back to the original value at higher voltages. The device with $d$ = 3.15~$\mu$m shows the $f_0$ drop of 12~MHz (corresponding to the relative decrease of 10.5~\% ) from 114~MHz at $V_P = 0$ down to 102~MHz for $V_P > $ 9~V. The devices with larger diameters show progressively larger strain-induced frequency downshift, i.e. 27~MHz for $d$ = 3.65~$\mu$m and 36~MHz for $d$ = 4.75~$\mu$m, corresponding to 27~\% and 45~\%, respectively. Fig.~\ref{fig3}(b) shows the $\Delta f$ vs. $d^2$ dependence with a quasi-linear increase of the frequency downshift with increased disk area.

To understand the observed size dependence, we calculated the $f_0$ values as a function of strain for different disk diameters [see Fig.~\ref{fig4}(a)]. Here, the same magnetic parameters were used as for the simulated data of Fig.~\ref{fig2}(b). For increased disk diameter, we observe a reduction of the critical strain, where the onset of the frequency $f_0$ decrease is observed. This suggests, that for larger disks, less magnetoelastic energy is needed to downshift the gyrotropic frequency. Note, that this effect does not depend on the magnetic anisotropy, and is present even for $K_u = 0$.

To further reveal the origin of the critical strain decrease for large disks, we calculated the equilibrium values of the magnetoelastic $W_{\varepsilon}$ and demagnetizing energies $W_d$ as a function of the compressive strain $\varepsilon_{xx}$ [see Fig.~\ref{fig4}(b,c)]. We exclude the exchange energy from the consideration due to its negligible value as compared to $W_{\varepsilon}$ and $W_d$ for the given disk sizes. One can see, that for small $\varepsilon_{xx}$ values, the magnetoelastic energy $W_{\varepsilon}$ dominates [see Fig.~\ref{fig4}(b)], however the onset of the $f_0$ decrease, for the given disk diameter, coincides with the onset of the demagnetizing energy $W_d$ increase. Fig.~\ref{fig4}(c) shows the $W_{\varepsilon} / W_d$ ratio vs. $\varepsilon_{xx}$ for different disk sizes. Here, the values of the critical strain (marked with a diamond) correspond to the maximum values of $W_{\varepsilon} / W_d$ i.e. to the onset of $W_d$ increase.
This suggests that besides the magnetoelastic energy contribution, the observed size effects may be related to the increased demagnetizing energy due to the strain-induced distortion of the magnetization distribution in the vortex.


\section{Conclusion}

We demonstrated a large frequency tunability (up to 45\%) in individual CoFeB disks accessed locally with low surface voltages ($\leqslant$16~V) applied to the PMN-PT substrate. Piezostrain-induced tuning and magnetoresistive readout of the VC dynamics via microwave rectification measurements allows for an all-electrical operation of the designed microdevices. The observed decrease of the VC gyrotropic frequency is associated with the strain-induced magnetoelastic energy. We showed that the frequency tunability strongly depends on the magnetic disk size, with increased frequency shift for the disks with larger diameter. Micromagnetic simulations show that the observed size effects originate jointly from the increased magnetoelastic energy as well as from the enhanced demagnetizing energy, resulting from the strain-induced distortion of the vortex configuration.
Our results show that electrically induced piezostrain offers an extra room for the frequency tunability of the VC dynamics in individual spintronic oscillators. In perspective, the frequency tunability for given strain magnitudes can be further enhanced by substituting CoFeB with large-magnetostriction materials (e.g. Fe$_{x}$Ga$_{1-x}$, TbFe$_2$, DyFe$_2$, Terfenol-D, etc).

\begin{acknowledgments}
Funded by the Deutsche Forschungsgemeinschaft (DFG, German Research Foundation) within the grant IU 5/2-1 (STUNNER) – project number 501377640.
Support from the Nanofabrication Facilities Rossendorf (NanoFaRo) at the IBC is gratefully acknowledged.
We thank Thomas Naumann for help with the CoFeB thin films growth.
We acknowledge useful discussions with Ciar\'an Fowley on the low-heat microfabrication process details.
We thank Ryszard Narkowicz for help with the implementation of the field modulation in the experimental setup.
\end{acknowledgments}

\bibliographystyle{apsrev4-1}
\bibliography{references}

\end{document}